%% file: ms.tex
\documentclass[apjpt4]{aastex}
\usepackage{psfig}

\newcommand{\Msun}{\mbox{\,$\rm{M_{\odot}}$}} 
\newcommand{\Rsun}{\mbox{\,$\rm{R_{\odot}}$}} 
\newcommand{\Lsun}{\mbox{\,$\rm{L_{\odot}}$}} 

\newcommand{\Xsun}{\mbox{\,$\rm{X_{\odot}}$}}

\newcommand{\Teff}{\mbox{\,$T_{eff}$}}
\newcommand{\Rstar}{\mbox{\,$R_{*}$}} 
 
\newcommand{\Minit}{\mbox{\,$M_{init}$}} 
\newcommand{\vturb}{\mbox{\,$v_{turb}$}}
\newcommand{\vrad}{\mbox{\,$v_{rad}$}}
\newcommand{\logg}{\mbox{\,$\log{g}$}}

\newcommand{\XO}{\mbox{\,$X_{O}$}}

\newcommand{\Htwo}{\mbox{\rm{H}$_2$}}
\newcommand{\HI}{\mbox{\rm{\ion{H}{1}}}}

\newcommand{\HeI}{\ion{He}{1}}
\newcommand{\HeII}{\ion{He}{2}}

\newcommand{\CIII}{\ion{C}{3}}
\newcommand{\CIV}{\ion{C}{4}}

\newcommand{\NIV}{\ion{N}{4}}
\newcommand{\NV}{\ion{N}{5}}

\newcommand{\OIV}{\ion{O}{4}}
\newcommand{\OV}{\ion{O}{5}}
\newcommand{\OVI}{\ion{O}{6}}

\newcommand{\eg}{\emph{e.g.}}
\newcommand{\ie}{\emph{i.e.}}
\newcommand{\doublet}{$\lambda\lambda$}
\newcommand{\singlet}{$\lambda$}

\newcommand{\EBMV}{\mbox{\,$E_{\rm{B-V}}$}}

\newcommand{\tlusty}{TLUSTY}
\newcommand{\synspec}{SYNSPEC}
\newcommand{\degree}[1]{\mbox{\,${#1}^o$}}

\newcommand{\Hbeta}{H$\beta$}

\newcommand{\Lyb}{Ly$\beta$}
\newcommand{\Lyg}{Ly$\gamma$}
\newcommand{\Lyd}{Ly$\delta$}

\newcommand{\Lyz}{Ly$\zeta$}

\newcommand{\kms}{\mbox{\,$\rm{km\:s^{-1}}$}}
\newcommand{\gunit}{\mbox{\,$\rm{cm\:s^{-2}}$}}

\def\star{Lo~1}

\begin{document}

\title{A FAR-UV SPECTROSCOPIC ANALYSIS OF THE CENTRAL STAR OF
  THE PLANETARY NEBULA LONGMORE~1\footnote{Based on observations made with
  the NASA-CNES-CSA Far Ultraviolet Spectroscopic Explorer and data
  from the MAST archive. FUSE is operated for NASA by the Johns
  Hopkins University under NASA contract NAS5-32985.}}

\author{J.E. Herald, L. Bianchi}
\vspace{1mm}
\affil{Department of Physics and Astronomy, The Johns Hopkins University}
\authoraddr{3400 N. Charles St., Baltimore, MD 21218-2411}
\email{herald@pha.jhu.edu,bianchi@pha.jhu.edu}
\received{2004 January 19}
\accepted{2004 February 20}

\begin{abstract}

We have performed a non-LTE spectroscopic analysis using far-UV and UV
data of the central star of the planetary nebula K1-26 (Longmore 1),
and found $\Teff = 120\pm10$~kK, $\logg =
6.7^{+0.3}_{-0.7}$~\gunit, and $y \simeq 0.10$.  The temperature is
significantly hotter than previous results based on optical line
analyses, highlighting the importance of analyzing the spectra of such
hot objects at shorter wavelengths.  The spectra show metal lines
(from, \eg, carbon, oxygen, sulfur, and iron).  The signatures of most
elements can be fit adequately using solar abundances, confirming the
classification of \star\ as a high gravity O(H) object.  Adopting a distance of
800~pc, we derive $\Rstar \simeq 0.04 $~\Rsun, $L \simeq 250 $~\Lsun,
and $M \simeq 0.6$~\Msun.  This places the object on the white dwarf
cooling sequence of the evolutionary tracks with an age of
$\tau_{evol} \simeq 65$~kyr.

\end{abstract}

\keywords{Planetary nebulae: individual (Longmore~1) --- stars:
  atmospheres --- stars: individual (Longmore 1) ---  stars: post-AGB
  --- stars: white dwarf   }

\section{INTRODUCTION}\label{sec:intro}

Longmore 1 (K1-26, PK 255-59 1, hereafter \star) was originally
discovered by \citet{longmore:77} as a PN having a notably large
angular size ($\sim400$\arcsec).  The spectra of its central star show
both hydrogen and \HeII\ absorption features, with no evidence of a
stellar wind in its UV or optical spectra \citep{patriarchi:91,
kaler:85, mendez:85}.  Because of its high galactic latitude ($b
\simeq -\degree{60}$), the reddening toward \star\ is thought to be minimal
\citep{kaler:85}.  Based on its optical spectrum, \citet{mendez:85}
termed \star\ an ``hgO(H)'' star - a high gravity object with very
broad Balmer absorptions.  Such objects can lie on the white-dwarf
cooling tracks, but can also be non-post-AGB objects.  A distance of
$D=800$~pc \citep{ishida:87} implies a nebular radius of $\sim0.8$~pc,
suggesting that \star\ is a quite evolved CSPN (most PN have radii
$\lesssim 0.5$~pc --- \citealp{cahn:92}).

Hot central stars of planetary nebulae (CSPN) emit most of their
observable flux in the Far-UV range.  We have observed the central
star of \star\ with the \emph{Far Ultraviolet Spectroscopic Explorer}
(FUSE) satellite in the 905---1187~\AA\ range.  Using this data as
well as archive \emph{International Ultraviolet Explorer} (IUE) data
(1150---3300~\AA), we determined the parameters of the central star
through stellar modeling, and discuss evolutionary implications.

Parameters of \star\ compiled from previous literature are listed in
Table~\ref{tab:lit_params}.

\section{OBSERVATIONS AND REDUCTION}\label{sec:obs}

Table~\ref{tab:obs} lists the spectra utilized in this paper.  \star\
was observed as part of FUSE's cycle 1 program P133 (Bianchi).  The IUE
data were retrieved from the MAST archive.  The observed spectra will
be presented in \S~\ref{sec:modeling}.

FUSE covers the wavelength range of 905--1187~\AA\ at a spectral
resolution of $\lesssim$ 30,000. The flux calibration accuracy of FUSE
is $\lesssim 10$~\% \citep{sahnow:00}.  It is described by
\citet{moos:00} and its on-orbit performance is discussed by
\citet{sahnow:00}. FUSE collects light concurrently in four different
channels (LiF1, LiF2, SiC1, and SiC2).  Each channel is recorded by
two detectors, each divided into two segments (A \& B) covering
different subsets of the above range with some overlap.

The FUSE spectra were taken through the LWRS
(30\arcsec$\times$30\arcsec) aperture.  These data, taken in
``time-tag'' mode, have been calibrated using the most recent FUSE
data reduction pipeline, efficiency curves and wavelength solutions
(CALFUSE v2.2). We combined the data from different segments, weighted
by detector sensitivity, and rebinning to a uniform dispersion of
0.05~\AA\ (which is probably close to the actual resolution since the
data were taken in the early part of the mission).  Bad areas of the
detectors, and those regimes affected by an instrumental artifact
known as ``the worm'' (FUSE Data Handbook v1.1), were excluded.  For
part of the first observation, telescope alignment problems moved the target
out of the LiF2/SiC2 aperture.  This also appears to have happened
with the SiC2 detector halfway through the second observation.  We
thus omitted data taken during these target drifts for the affected
detectors.

Four IUE spectra of the central star of \star\ are available, however
it appears that one (SWP20275) missed the central star.  The only
high-resolution spectrum (SWP39146) is under-exposed and was not
used.  The two remaining low resolution long wavelength and short
wavelength spectra are in agreement in the region of overlap.  The IUE
spectra are relatively featureless and are mainly used to fit the
continuum flux distribution.

\section{MODELING}\label{sec:modeling}

\begin{figure}[htbp]
\begin{center}
\epsscale{.3}
\rotatebox{270}{\plotone{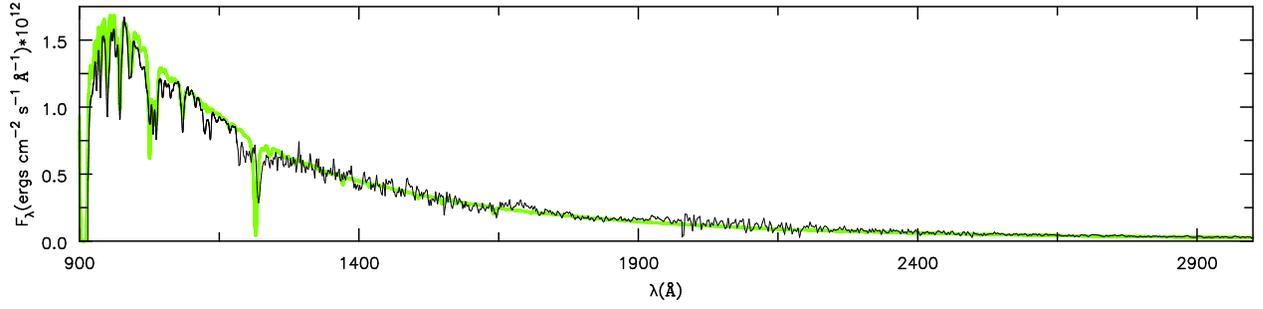}}
\caption{FUV-UV (FUSE and IUE) spectra of \star\ are shown (black)
  along with our stellar models (red/dark gray) with our
  hydrogen absorption model applied (convolved with a 3~\AA\ Gaussian).
  The model, with no reddening applied, does a good job
  at matching the observed flux distribution.
  }
\label{fig:uv}
\end{center}
\end{figure}

Modeling of \star\ consisted of two parts: modeling the hot central
star, and modeling the sight-line hydrogen (atomic and molecular).  We
describe each in turn.  A virtue of its location significantly
outside the Galactic plane is a low reddening, which we determine, by
fitting the continuum slope, to be $\EBMV < 0.01$ (we use a value of $\EBMV = 0$
throughout this paper).  The data, as well as the
model fits, are shown in Figs.~\ref{fig:fuv} and \ref{fig:uv}.  We
determine the radial velocity of \star\ to be $\vrad = 100 \pm
10$~\kms\ using stellar absorption line features in the long
wavelength FUSE range ($>1050$~\AA) such as \CIV\ \singlet 1169.0 and
\CIV\ \singlet 1107.6.

\subsection{The Central Star Model}\label{sec:wd}

To model the white dwarf central star, we used the \tlusty\ code to
calculate the stellar atmosphere, and \synspec\ to calculate the
synthetic flux \citep{hubeny:88, hubeny:92, hubeny:94, hubeny:95}.  \tlusty\
calculates the atmospheric structure assuming radiative and
hydrostatic equilibrium, and a plane-parallel geometry, in
non-LTE (NLTE) conditions.  In the case of hot (\Teff
$\gtrsim$ 50~kK) white dwarfs, LTE calculations are not appropriate,
and result in significant deviations from NLTE calculations
(see, \eg, \citealp{dreizler:96, werner:96, werner:91,
napiwotzki:97}). This is because in such hot objects, the populations of
the ions are mainly determined by the intense radiation field despite
the high gravities. 

The FUSE spectrum shows features of hydrogen, helium, and metals (\eg,
C, O, Fe, and S), with \OVI\ \doublet 1032,38 being especially
prominent.  Test models indicated that the solar abundance ratio for
H/He (as found by \citealp{mendez:85}) was adequate.  We thus
constructed a grid of solar abundance models varying \Teff\ and \logg,
treating hydrogen and helium in NLTE to calculate the structure of the
atmosphere.  Once \Teff\ and \logg\ were determined adequately, they
were held fixed, and the CNO elements were varied
individually in NLTE to constrain their abundances and ensure that
neglect of their NLTE treatment did not alter the derived \Teff\ and
\logg.

The atomic data used come from TOPBASE, the data-base of the Opacity
Project \citep{cunto:93}.  \tlusty\ makes use of the concept of
``superlevels'', where levels of similar energy are grouped together
and treated as a single level in the rate equations (after
\citealp{anderson:89}).  The number of levels+superlevels used for the
NLTE model ions were: \HI(8+1), \HeI (24+0), \HeII (20+0), \CIII (34+12),
\CIV (35+2), \NIV(15+8), \NV(21+4), \OIV (39+31), \OV (34+6), and \OVI
(15+5).  Ne, Na, Mg, Si, S, Ar, Ca, and Fe were allowed to contribute
to the total number of particles and charge but their opacity
contribution was neglected in the model atmosphere calculation.  We
have adopted \synspec's values for solar abundances, which are taken
from \citet{grevesse:98}.

\begin{figure}[htbp]
\begin{center}
\epsscale{.3}
\rotatebox{270}{\plotone{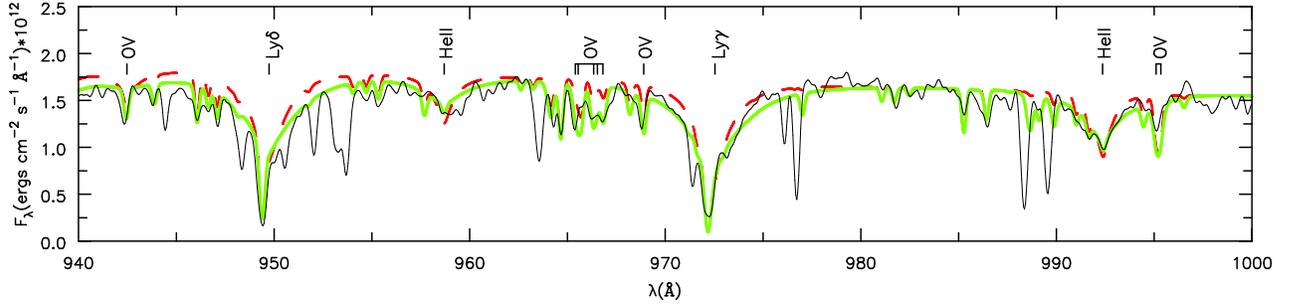}}
\caption{Constraining the gravity: A portion of the FUSE spectrum is
  shown (black), along with our stellar model ($\Teff = 120$~kK) with
  $\logg=6.0$~\gunit\ (dashed red/dark gray) and $\logg=7.0$~\gunit\ (green/light
  gray).  Based on the wings of the \Lyd, \Lyg, and the \HeII\
  features, the gravity lies between these two values.  We derive
  $\logg = 6.7^{+0.3}_{-0.7}$~\gunit. }
\label{fig:logg}
\end{center}
\end{figure}

As previously mentioned, a solar hydrogen to helium ratio appeared
adequate to fit the \HeII\ and H Lyman spectrum.
The gravity was constrained by fitting mainly the wings of these features, as
demonstrated in Fig.~\ref{fig:logg}.  To determine the effective
temperature, the FUV-UV continuum shape, as well as FUV spectral
features of hydrogen, helium and metals (mainly oxygen and carbon)
were used as diagnostics.  The parameters of our ``best'' fit model for the
CSPN of \star\ are: $\Teff=120\pm10$~kK,
$\logg=6.7^{+0.3}_{-0.7}$~\gunit.  Solar values for the metal
abundances were found to be adequate, except for oxygen, for which the
solar value underproduced the strong \OVI\ \doublet 1032,38 feature
(shown in Fig.~\ref{fig:OVI}).  We found that an oxygen abundance enriched
5 times with respect to the solar value ($\XO = 5 \Xsun$ by mass)
produced a good fit, however some of the other oxygen features then
appeared a bit strong.  For temperatures $\Teff \simeq 100-120~$~kK,
the \OVI\ feature is at its strongest, thus higher or lower
temperatures require an even greater oxygen enrichment.
Unless otherwise stated, the model spectrum shown in the
figures and what we refer to as ``our model'' has the parameters
$\Teff=120$~kK, $\logg = 6.7$~\gunit, $\XO=5\Xsun$, with the
abundances of all other elements set to their solar values.

\subsection{Modeling \Htwo\ and \HI\ absorption toward \star}\label{sec:htwo}

\begin{figure}[htbp]
\begin{center}
\epsscale{0.70}
\rotatebox{0}{\plotone{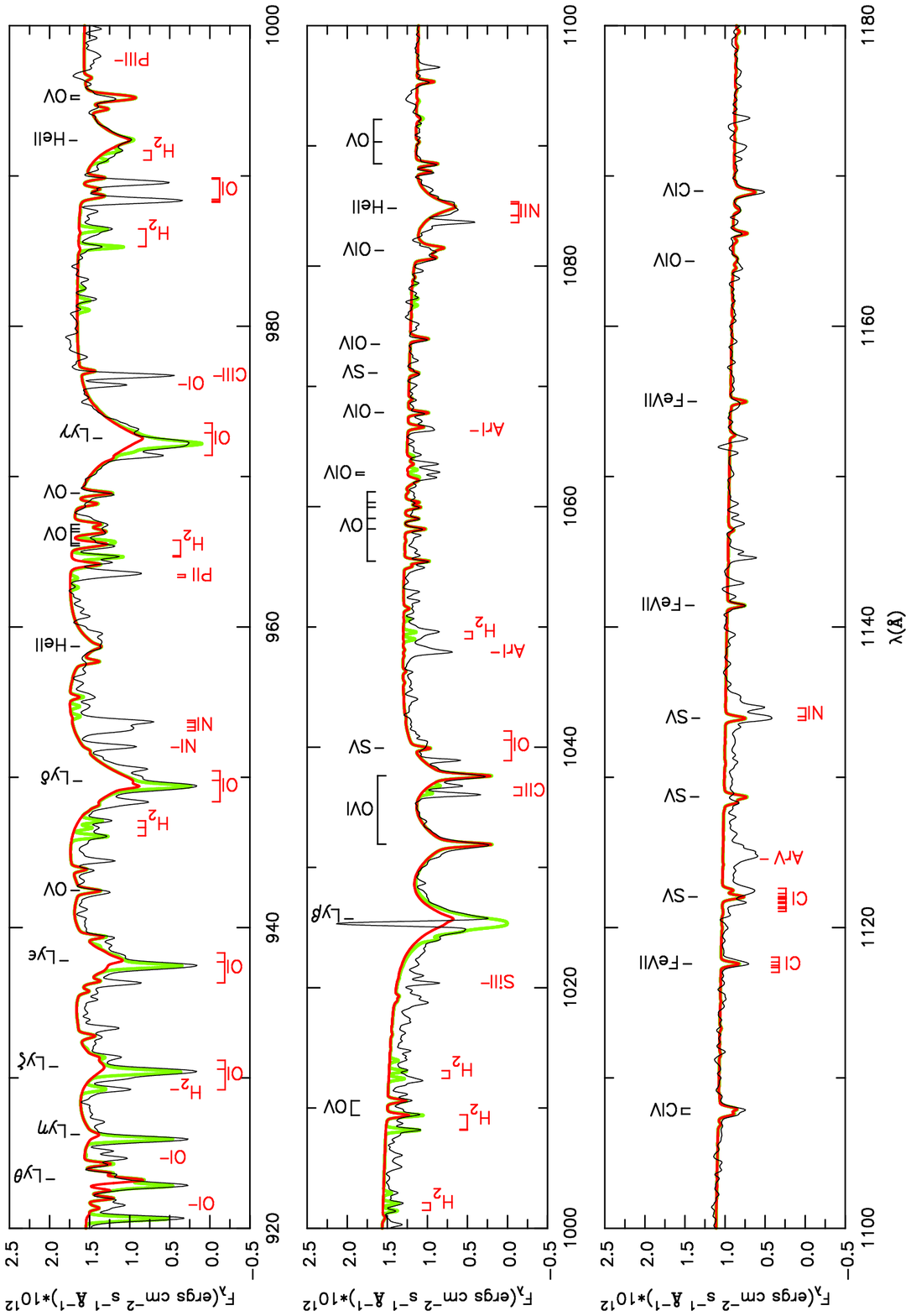}}
\caption{FUV: The FUSE spectrum of \star\ is shown (black) along with
  our stellar model (described in \S~\ref{sec:wd}), both with (green/light gray) and without
  (red/dark gray) our hydrogen absorption model applied.  Prominent
  stellar features are marked with black labels, interstellar
  absorption features and airglow lines with red/gray labels. The
  Lyman features consist of a broader, stellar component, and a
  narrower, interstellar component, separated by $\simeq 100$~\kms\ in
  velocity space (the \Lyb\ emission is airglow).  All spectra have been
  convolved with a Gaussian of 0.25~\AA\ for clarity.  Virtually all
  the stellar and hydrogen features are well-matched -- most
  non-matched features are of interstellar origin.}
\label{fig:fuv}
\end{center}
\end{figure}

\begin{figure}[htbp]
\begin{center}
\epsscale{.5}
\rotatebox{270}{\plotone{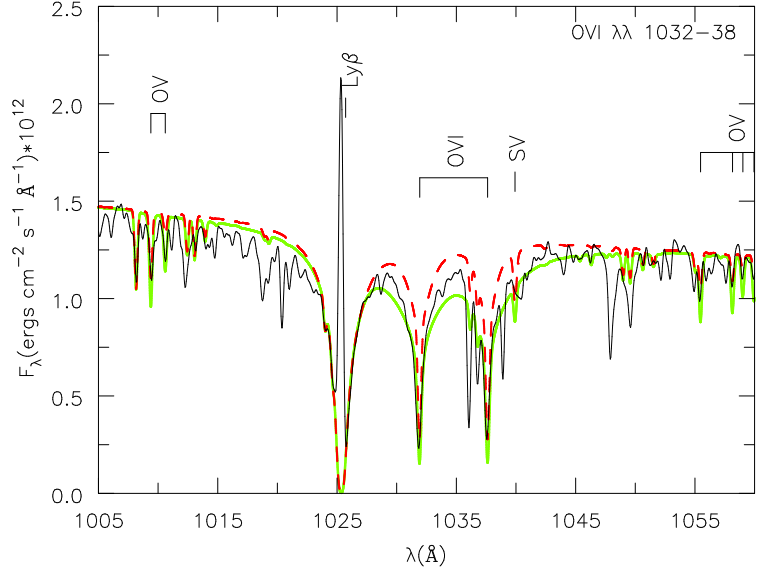}}
\caption{Constraining the oxygen abundance: The FUSE spectrum in the
  region of \OVI\ \doublet 1032,38 (black) is shown, along with a
  stellar model with solar oxygen abundance (dashed red/dark gray) and
  a model with oxygen enriched 10 times with respect to the solar
  value (green/light gray).  The former underproduces the \OVI\
  doublet, while the latter overproduces the feature.  Our final
  model, with an oxygen abundance of 5 times the solar value, fits the
  the \OVI\ doublet well (shown in Fig.~\ref{fig:fuv}).  }
\label{fig:OVI}
\end{center}
\end{figure}

The FUSE spectrum of \star\ (Fig.~\ref{fig:fuv}) displays a series of 
absorption features corresponding to the hydrogen Lyman sequence.  The
cores of these features are attributable to absorption from sight-line
hydrogen.  These cores are velocity shifted with respect to the
broader, stellar Lyman absorption features by $\simeq 100$~\kms, which
corresponds to our measured radial velocity for the CSPN lines.
Thus these features are interstellar in origin (rather than
circumstellar).

The effects of \HI\ absorption were applied to the model
spectrum in the following manner.  For a given column density ($N$)
and gas temperature ($T$), the absorption profile of each line is
calculated by multiplying the line core optical depth ($\tau_0$) by
the Voigt profile $[H(a,x)]$ where $x$ is the frequency in Doppler
units and $a$ is the ratio of the line damping constant to the Doppler
width (the ``b'' parameter).  The observed flux is then $F_{obs} =
 \exp{[-\tau_{0}H(a,x)]} \times F_{intrinsic}$.

Because the \HI\ column density determination is insensitive to
temperature, we determine $N(\HI)$ by assuming $T(\HI)=80$~K
(corresponding to the mean temperature of the ISM ---
\citealp{hughes:71}) and $\vturb = 10$~\kms\ and fitting the Lyman
profiles of the FUSE data.  Doing so, we derive
$\log{N(\HI)}=20.3^{+0.4}_{-0.3}$~cm$^{-2}$.

The FUSE spectrum also shows some weak absorption features from
intervening molecular hydrogen, which originate from the Lyman ($B^1
\Sigma^+_u$--$X^1 \Sigma^+_g$) and Werner ($C^1 \Pi^{\pm}_u$--$X^1
\Sigma^+_g$) sequences (these are marked in Fig.~\ref{fig:fuv}).  We
applied the effects of different \Htwo\ models in a similar manner as
the \HI, again assuming a gas temperature of 80~K.  We derive a
relatively small column density of
$\log{N(\Htwo)}=14.9\pm0.2$~cm$^{-2}$.  The low column density is
probably a consequence of the high galactic latitude of \star, and is
consistent with our determination of $\EBMV < 0.01$~mag, based on
typical relations between \EBMV\ and \Htwo\ column densities in the
ISM found by \citet{bohlin:78}.

Our stellar model spectrum, with hydrogen absorptions corresponding
to $\log{N(\HI)}=20.3$~cm$^{-2}$ and
$\log{N(\Htwo)}=14.9$~cm$^{-2}$ applied, is shown in
Fig.~\ref{fig:fuv}.

\section{DISCUSSION}\label{sec:discussion}

Our derived model parameters for the central star and sight-line
hydrogen are presented in Table~\ref{tab:wd_param}.  Scaling our model
flux to the observed flux yields $\Rstar/D$, the ratio of the stellar
radius to the distance.  This value, using a distance of $D=800$~pc
\citep{ishida:87}, yields a radius of $\Rstar \simeq 0.037$~\Rsun\ and
a corresponding luminosity of $L \simeq 250$~\Lsun.  The model flux
then yields a corresponding visual magnitude of $V = 15.5$~mag, in
good agreement with the measured value of $V = 15.4$~mag
\citep{kaler:85}.  Because we can only constrain the gravity rather
loosely, we cannot derive a meaningful value for the mass of the
central star without appealing to stellar evolution tracks.  As
discussed in \S~\ref{sec:wd}, the \OVI\ doublet may indicate an oxygen
enriched atmosphere.  Usually, in CSPN, oxygen enrichment is associated
with helium-rich objects (\ie, helium-burners), and is often
accompanied by carbon-enrichment.  However, the abundances of the
other elements in \star\ do not appear to be much different than the
solar values, which is characteristic of many H-burning CSPN.
Therefore, we compared our derived effective temperature and
luminosity with both the hydrogen- and helium-burning (solar abundance)
tracks of \citet{vassiliadis:94}.  The H-burning tracks indicate a
current core mass of $M_c = 0.633$~\Msun, and an initial mass of 2.0~\Msun,
with the uncertainties of our parameters encompassing the
(\Minit,$M_c$) = (1.5,0.597) and (2.5,0.677) tracks as well.  So, from
the H-burning evolutionary models, we derive (\Minit,$M_c$) =
($2.0\pm0.5$,$0.63\pm0.04$)~\Msun\ and a post-AGB age
$\tau_{evol}\sim$60~kyr.  In a similar fashion, we derive
(\Minit,$M_c$) = ($1.5\pm0.5$,$0.60\pm0.04$)~\Msun,
$\tau_{evol}\sim$70~kyr from comparison with the He-burning tracks.

The derived stellar parameters of \star\ are similar to those of the
hotter, higher-gravity O(H) stars in the sample of central stars for
old PN classified by \citet{napiwotzki:99}.  Thus, we confirm the
\citet{mendez:85} classification of \star\ as a hgO(H) star.

\citet{mendez:85} performed a non-LTE analysis of its optical spectrum
and obtained $\Teff = 65\pm10$~kK, $\logg = 5.7\pm0.3$~\gunit, and
$y=0.10\pm0.03$ for the central star.  We have calculated a TLUSTY
model with these parameters and find it fails to match the FUV data
for multiple reasons.  When this model is scaled to match the UV
continuum level, it significantly underproduces the FUV continuum
level.  It also fails to duplicate many of the FUV diagnostics, most
notably the strong \OVI\ \doublet 1032,38 feature.  Similarly,
\citet{hoare:96}, from an analysis of the optical and extreme
ultraviolet spectra of the CSPN NGC~1360, found a temperature
significantly higher than the results of \citet{mendez:85}, which were
based on optical data only.  Our significantly higher derived
temperature and gravity illustrate the importance of considering the
FUV wavelength regime when modeling such hot CSPN, where they emit the
majority of their observable flux (\ie, longwards of the Lyman limit)
as well as display their strongest stellar features.

\section{CONCLUSIONS}\label{sec:conclusions}

We have analyzed FUV and UV spectra of \star, a hot CSPN notable for
its relatively high galactic latitude and thus having a minimal reddening.
Its FUSE spectrum, aside from showing hydrogen and helium lines, shows
strong \OVI\ \doublet 1032,38 signatures, perhaps indicating an
oxygen-enriched object.  

We have modeled the FUSE and IUE spectrum of this object to determine
parameters of $\Teff = 120$~kK, $\logg = 6.7$~\gunit, $\Rstar =
0.04$~\Rsun, $L = 250$~\Lsun, and $M \simeq 0.6$~\Msun.  The
temperature is much higher than that derived by \citet{mendez:85} from
an optical-line analysis ($\Teff = 65\pm10$~kK), and illustrates the
importance of the FUV-UV range in the analysis of hot CSPN.  These
parameters confirm the \citet{mendez:85} classification of \star\ as a
high-gravity O(H) star.  Comparison of our parameters to evolutionary
tracks indicate a post-AGB age of $\sim 65$~kyr.  We also measure
$\vrad \simeq 100$~\kms\ for the \star\ PN system.

\acknowledgements 

We thank Terry Lanz and Ivan Hubeny for their help with the \tlusty\
code.  We thank Stephan McCandliss for making his \Htwo\ molecular
data and tools available.  We also are grateful to the referee, Klaus
Werner, for his constructive comments.  The SIMBAD database was used
for literature searches.  This work has been funded by NASA grant NAG
5-9219 (NRA-99-01-LTSA-029).  The IUE data presented in this paper
were obtained from the Multimission Archive (MAST) at the Space
Telescope Science Institute (STScI). STScI is operated by the
Association of Universities for Research in Astronomy, Inc., under
NASA contract NAS5-26555. Support for MAST for non-HST data is
provided by the NASA Office of Space Science via grant NAG5-7584 and
by other grants and contracts.



\newpage

\clearpage

\clearpage

\clearpage

\clearpage


\newpage
\input{tb1}
\newpage
\input{tb2}
\newpage
\input{tb3}

\end{document}

%% file: tb1.tex
\begin{table}[htb]
\scriptsize
\caption{Parameters of \star}\label{tab:lit_params}
\begin{tabular}{lcl}
\hline
Quantity & Value & References (Notes) \\
\hline
R.A. (J2000) & 02 56 58.23 & \\
Dec. (J2000) & --40 10 19.41 & \\
Galactic latitude ($b$) & --59.64 & \citet{ishida:87} \\
Galactic longitude ($l$) & 255.35 & \citet{ishida:87} \\
Height ($z$) [pc] & --690 & \citet{ishida:87} \\
Lateral distance ($q$) [pc] & 400 & \citet{ishida:87}\\
Radial distance $D$ [pc] & 800 & \citet{ishida:87}\\
PN radius $R_{PN}$ [\arcsec] & 187, 230$\times$192 &\citet{longmore:77}, \citet{kohoutek:77}\\
PN radius $R_{PN}$ [pc] & 0.72, 0.89$\times$0.75 & Assuming $D=800$~pc\\
$v_{rad}$ [kms] & $65\pm21$ & \citet{west:85} \\
Logaritmic extinction at \Hbeta\ ($c$) & $0.0\pm0.05$ & \citet{kaler:85b} \\
CSPN $V$ [mag] & 15.4 & \citet{kaler:85} \\ 
CSPN \Teff\ [kK] & $65\pm10$ & \citet{mendez:85}, from optical analysis\\
CSPN \logg & $5.7\pm0.3$ & \citet{mendez:85}, from optical analysis \\
CSPN He/H  & $0.1\pm0.03$ &  \citet{mendez:85}, from optical analysis \\
\hline
\end{tabular}
\end{table}

%% file: tb2.tex
\begin{table}[htbp]
\caption{\star: Utilized Spectra}\label{tab:obs}
\begin{tabular}{ccccc}
\hline
Instrument & Data & Date & Resolution & Aperture \\
           &  Set &      & (\AA) & (\arcsec) \\
\hline
FUSE & P1330601001 & 12/11/00 & $\sim0.05$ & $30\times30$ \\
FUSE & P1330601002 & 12/11/00 & $\sim0.05$ & $30\times30$ \\
IUE  & LWP11432 & 8/19/87 & 5-6 & $10\times20$\\
IUE  & SWP21421 & 11/1/83 & 5-6 & $10\times20$\\
\hline
\end{tabular}
\end{table}

%% file: tb3.tex
\begin{table}[htbp]
\caption{Derived Parameters for \star}\label{tab:wd_param}
\begin{tabular}{lcl}
\hline
Parameter & Value & Primary Diagnostics and Comments \\
\hline
CSPN \Teff\ [kK] & 120$\pm$10 & H~\Lyz--\Lyb, \HeII, metal features\\
CSPN \logg\ [cm s$^{-2}$] & $6.70^{+0.3}_{-0.4}$ & H~\Lyz--\Lyb, \HeII\ wings\\
CSPN $\Rstar/D$ [\Rsun/pc] & $(4.6\pm0.3)\times10^{-5}$ & Scaling
model flux to observed UV flux \\
CSPN \XO\ [\Xsun] & $1$--$5$ &\OVI\ \doublet 1032,38, other oxygen features  \\
CSPN $\Rstar$ [\Rsun] & $(3.7\pm0.2)\times10^{-2}$ & Using $D=800$~pc \\
CSPN $L$ [\Lsun] & $250^{+140}_{-100}$ & - \\
\EBMV\ [mag] & $<$0.01 & Continuum shape\\
I.S. $\log{N(\HI)}$ [cm$^{-2}$] & $20.3^{+0.4}_{-0.3}$ & Lyman features, $T=80$~K assumed \\
I.S. $\log{N(\Htwo)}$ [cm$^{-2}$] & $14.9\pm0.2$ &  FUV \Htwo\ features, $T=80$~K assumed\\
$V_{\star}$ [\kms] & $-100\pm10$ & photospheric absorption lines\\
\hline
\end{tabular}
\end{table}

%% file: ms.bbl
\begin{thebibliography}{27}
\expandafter\ifx\csname natexlab\endcsname\relax\def\natexlab#1{#1}\fi

\bibitem[{Anderson(1989)}]{anderson:89}
Anderson, L.~S. 1989, \apj, 339, 588

\bibitem[{Bohlin {et~al.}(1978)Bohlin, Savage, \& Drake}]{bohlin:78}
Bohlin, R.~C., Savage, B.~D., \& Drake, J.~F. 1978, \apj, 224, 132

\bibitem[{Cahn {et~al.}(1992)Cahn, Kaler, \& Stanghellini}]{cahn:92}
Cahn, J.~H., Kaler, J.~B., \& Stanghellini, L. 1992, \aaps, 94, 399

\bibitem[{Cunto {et~al.}(1993)Cunto, Mendoza, Ochsenbein, \&
  Zeippen}]{cunto:93}
Cunto, W., Mendoza, C., Ochsenbein, F., \& Zeippen, C.~J. 1993, \aap, 275, 5

\bibitem[{Dreizler \& Werner(1996)}]{dreizler:96}
Dreizler, S. \& Werner, K. 1996, \aap, 314, 217

\bibitem[{Grevesse \& Sauval(1998)}]{grevesse:98}
Grevesse, N. \& Sauval, A.~J. 1998, \ssr, 85, 161

\bibitem[{Hoare {et~al.}(1996)Hoare, Drake, Werner, \& Dreizler}]{hoare:96}
Hoare, M.~G., Drake, J.~J., Werner, K., \& Dreizler, S. 1996, \mnras, 283, 830

\bibitem[{Hubeny(1988)}]{hubeny:88}
Hubeny, I. 1988, Comput. Phys. Comm., 52, 103

\bibitem[{Hubeny {et~al.}(1994)Hubeny, Hummer, \& Lanz}]{hubeny:94}
Hubeny, I., Hummer, D.~G., \& Lanz, T. 1994, \aap, 282, 157

\bibitem[{Hubeny \& Lanz(1992)}]{hubeny:92}
Hubeny, I. \& Lanz, T. 1992, \aap, 262, 501

\bibitem[{Hubeny \& Lanz(1995)}]{hubeny:95}
---. 1995, \apj, 493, 875

\bibitem[{Hughes {et~al.}(1971)Hughes, Thompson, \& Colvin}]{hughes:71}
Hughes, M.~P., Thompson, A.~R., \& Colvin, R.~S. 1971, \apjs, 23, 323

\bibitem[{Ishida \& Weinberger(1987)}]{ishida:87}
Ishida, K. \& Weinberger, R. 1987, \aap, 178, 227

\bibitem[{Kaler \& Feibelman(1985)}]{kaler:85b}
Kaler, J.~B. \& Feibelman, W.~A. 1985, \apj, 297, 724

\bibitem[{Kaler \& Lutz(1985)}]{kaler:85}
Kaler, J.~B. \& Lutz, J.~H. 1985, \pasp, 97, 700

\bibitem[{Kohoutek \& Laustsen(1977)}]{kohoutek:77}
Kohoutek, L. \& Laustsen, S. 1977, \aap, 61, 761

\bibitem[{Longmore(1977)}]{longmore:77}
Longmore, A.~J. 1977, \mnras, 178, 251

\bibitem[{M\`{e}ndez {et~al.}(1985)M\`{e}ndez, Kudritzki, \& Simon}]{mendez:85}
M\`{e}ndez, R.~H., Kudritzki, R.~P., \& Simon, K.~P. 1985, \aap, 142, 289

\bibitem[{Moos {et~al.}(2000)Moos, Cash, \& Cowie}]{moos:00}
Moos, H.~W., Cash, W.~C., \& Cowie, L.~L. 2000, \apj, 538, 1

\bibitem[{Napiwotzki(1997)}]{napiwotzki:97}
Napiwotzki, R. 1997, \aap, 322, 256

\bibitem[{Napiwotzki(1999)}]{napiwotzki:99}
---. 1999, \aap, 350, 101

\bibitem[{Patriarchi \& Perinotto(1991)}]{patriarchi:91}
Patriarchi, P. \& Perinotto, M. 1991, \aaps, 91, 325

\bibitem[{Sahnow {et~al.}(2000)Sahnow, Moos, \& Ake}]{sahnow:00}
Sahnow, D.~J., Moos, M.~W., \& Ake, T.~B. 2000, \apj, 538, 7

\bibitem[{Vassiliadis \& Wood(1994)}]{vassiliadis:94}
Vassiliadis, E. \& Wood, P.~R. 1994, \apj, 92, 125

\bibitem[{Werner(1996)}]{werner:96}
Werner, K. 1996, \aap, 309, 861

\bibitem[{Werner {et~al.}(1991)Werner, Heber, \& Hunger}]{werner:91}
Werner, K., Heber, U., \& Hunger, K. 1991, \aap, 244, 437

\bibitem[{West \& Kohoutek(1985)}]{west:85}
West, R.~M. \& Kohoutek, L. 1985, \apjs, 60, 91

\end{thebibliography}
